\documentclass[12pt,a4paper,superscriptaddress,aps,prd,nofootinbib,notitlepage]{revtex4-1}
\usepackage{lmodern}

\usepackage[T1,T5]{fontenc}
\usepackage[utf8]{inputenc}
\usepackage{color}
\usepackage{amsmath}
\usepackage{amssymb}
\usepackage{graphicx}
\usepackage{graphics}
\usepackage{slashed}
\usepackage{tikz-feynman}
\usepackage[stable]{footmisc}
\usepackage{natbib}

\makeatletter

\pdfpageheight\paperheight
\pdfpagewidth\paperwidth

\definecolor{somegreen}{cmyk}{0,0.49,0.98,0.09}
\definecolor{red}{rgb}{1,0,0}
\definecolor{magenta}{cmyk}{0,1,0,0}
\definecolor{violet}{cmyk}{0,1,0,0}
\definecolor{darkgreen}{rgb}{0,0.65,0.05}
\definecolor{antiquefuchsia}{rgb}{0.33, 0.1, 0.89}

\makeatother

\newcommand{\be}{\begin{equation}}
\newcommand{\ee}{\end{equation}}
\newcommand{\en}{\end{equation}}
\newcommand{\ba}{\begin{eqnarray}}
\newcommand{\ea}{\end{eqnarray}}
\newcommand{\bea}{\begin{eqnarray}}
\newcommand{\eea}{\end{eqnarray}}
\newcommand{\pa}{\partial}

\def\As{A\!\!\!/}
\def\ks{k\!\!\!/}

\def\ps{p\!\!\!/}

\def\bs{b\!\!\!/}

\def\ds{\partial\!\!\!/}

\begin{document}


\title{Non-Abelian aether-like term in four dimensions}
 \author{A. J. G. Carvalho}
\affiliation{Departamento de F\'{\i}sica, Universidade Federal da Para\'{\i}ba\\
 Caixa Postal 5008, 58051-970, Jo\~ao Pessoa, Para\'{\i}ba, Brazil}
\email{jroberto, petrov@fisica.ufpb.br, gomescarvalhoantoniojose@gmail.com}
 \author{D. R. Granado}
\affiliation{
Duy T\^{a}n University, Institute of Research and Development, \\
 P809, 3 Quang Trung, H\h{a}i Ch\^{a}u, \DJ\`{a} N\~\abreve ng, Vietnam} 
 \email{diegorochagranado@gmail.com}
\author{J. R. Nascimento}
\affiliation{Departamento de F\'{\i}sica, Universidade Federal da Para\'{\i}ba\\
 Caixa Postal 5008, 58051-970, Jo\~ao Pessoa, Para\'{\i}ba, Brazil}
\email{jroberto, petrov@fisica.ufpb.br, gomescarvalhoantoniojose@gmail.com}
\author{A. Yu. Petrov}
\affiliation{Departamento de F\'{\i}sica, Universidade Federal da Para\'{\i}ba\\
 Caixa Postal 5008, 58051-970, Jo\~ao Pessoa, Para\'{\i}ba, Brazil}
\email{jroberto, petrov@fisica.ufpb.br, gomescarvalhoantoniojose@gmail.com}

\begin{abstract}
The non-Abelian aether-like Lorentz-breaking term, involving triple and quartic self-coupling vertices, is generated from the non-Abelian generalization of the Lorentz-breaking extended QED including only a minimal spinor-vector interaction. This term is shown explicitly to be finite and non-ambiguous.
\end{abstract}

\maketitle

\section{Introduction}

The Lorentz symmetry breaking opens broad possibilities for constructing extensions of known field theory models. The first steps in this direction were presented in the paradigmatic papers by Kostelecky and Colladay \cite{ColKost1,ColKost2}. In these papers a list of possible Lorentz-breaking extensions of the standard model has been presented for the first time. 
 Further, many terms of this list were shown to arise as one-loop perturbative corrections. In the case of scalar, gauge and gravitational fields such terms arise from the corresponding fermionic determinants (for a review of various situations where such arising occurs, see \cite{ourreview}).

However, absolute majority of these results describe Lorentz-breaking contributions of second order in fields -- for example, Carroll-Field-Jackiw (CFJ) term, aether terms for scalar and gauge fields, higher-derivative contributions for gauge fields. At the same time, it is well known that many phenomenological interesting results are obtained in essentially non-Abelian gauge theories (the most interesting application of these theories consists certainly in studies of QCD and confinement) whose Lagrangian involves terms up to fourth order in fields. 

The first example of a non-Abelian Lorentz-breaking term is the non-Abelian CFJ term \cite{ColMac,YMCS}  which breaks the Lorentz symmetry but preserves the gauge symmetry. In \cite{ouraether}, the authors presented the appropriate scheme for the  path integral quantization of Yang-Mills-CFJ system. In order to  remove properly the gauge copies, one must restrict the path integral to a subspace of independent dynamical fields called Gribov region. In \cite{Gribov:1977wm}, it was shown for the first time in the YM system that, in order to deal with the gauge copies, such restriction is mandatory and as a consequence implies the modification of the gluon propagator. In the low-energy limit of the theory, the gluon propagator exhibits the propagation of non-physical particles, which means that we are no longer able to describe the propagation of the real degrees of freedom of the theory in its low-energy limit. This feature is interpreted as the description of the confinement problem of gluons, which is an inherent problem of non-Abelian gauge theories. In \cite{ouraether}, a new non-Abelian Lorentz-breaking model, that is, the non-Abelian aether-like model was proposed, a  simplified scheme for the generation of the non-Abelian aether-like term was presented and the proper path integral quantization of the Yang-Mills-aether-like system was treated. By means of the Gribov description of the gluon confinement problem mentioned above, in \cite{ourCFJ,ouraether}, it was verified whether the non-Abelian Lorentz symmetry breaking terms can influence the theory in the low-energy limit in a manner implying that the confinement problem may not occur for certain values of the coupling constants presented in the theory. In both papers it was shown that due to the small value of the Lorentz-breaking parameters ($\approx10^{-7}$ GeV), these terms does not affect the confinement regimes of the theory.  Therefore, in a full analogy with those papers, it is natural to expect that the term generated by us here can yield only very tiny modifications within the confinement scenario.
 
Thus the natural problem consists of studying the various issues related to non-Abelian Lorentz-breaking terms. Some examples of such terms were recently listed in \cite{KosLi}. However, up to now, there are only very few studies of such terms -- as it was mentioned before, the non-Abelian extension of the CFJ term was considered in  \cite{ColMac,YMCS} and some non-perturbative effects for Lorentz-breaking extensions of Yang-Mills theory were discussed in \cite{Sobr}. Therefore, it is natural to look for more new results in this direction. The most natural problem from the perturbative generation viewpoint could be the generation of the non-Abelian aether term using only minimal couplings. As it is known, this generation makes the result in the Abelian case to be superficially finite and non-ambiguous \cite{Scarp}, avoiding the problem of regularization dependence. Namely, this scheme of calculations will be generalized to the non-Abelian case, i.e., we will obtain CPT-even three and four-point functions of the gauge field. 

This paper is organized as follows. In the section 2, we give basic definitions; in the section 3, we perform the one-loop calculations; finally, in the section 4 we present our results. In the Appendix, the relevant momentum integrals are given.

\section{The non-Abelian aether term}

Now, let us start with discussion of the non-Abelian aether term. The original aether term \cite{Carroll} is known to have the following form
\begin{equation}
\label{aetherlagr}
{\cal L}_{aether}= u^{\mu}u_{\nu}F_{\mu\lambda}F^{\nu\lambda},
\end{equation}
where $u_{\mu}$ is a some constant vector, and $F_{\mu\nu}$ is the usual stress tensor of the electromagnetic field. Actually, this term is nothing more as the general CPT-even term
\begin{equation}
{\cal L}_{even}= \kappa^{\mu\nu\lambda\rho}F_{\mu\nu}F_{\lambda\rho},
\end{equation}
proposed in \cite{ColKost1,ColKost2}, for a special form of the constant tensor $\kappa^{\mu\nu\lambda\rho}$ (for some issues related to this CPT-even term, including its impacts for the plane wave solutions of modified Maxwell equations, see also f.e. \cite{Ferr}).
It is clear that the non-Abelian analogue of the term (\ref{aetherlagr})  can be written down straightforwardly -- it is sufficient to replace the Abelian stress tensor by its non-Abelian analogue \cite{ouraether}, i.e.
\begin{equation}
\label{aetherlagr1}
{\cal L}_{aether,YM}= u^{\mu}u_{\nu}{\rm tr}(F_{\mu\lambda}F^{\nu\lambda}),
\end{equation}
where $F_{\mu\nu}=F_{\mu\nu}^aT^a$ is the non-Abelian, Lie-algebra valued stress tensor.

In \cite{Gomes:2009ch} the scheme for generation of the Abelian aether term was proposed, and this term was obtained as a one-loop quantum correction in a theory which involves a magnetic coupling of the fermion to an electromagnetic field. The generalization of the scheme used in \cite{Gomes:2009ch}, to the non-Abelian case is straightforward. In order to, one can start with the following classical action:
\begin{equation}
S_{\psi}=\int d^4x\sum_{i,j=1}^N\bar{\psi}_i(i\delta_{ij}\ds-g^{\prime}\epsilon^{\mu\nu\lambda\rho}F^a_{\mu\nu}b_{\lambda}\gamma_{\rho}(T^a)_{ij}-m\delta_{ij})\psi_j.
\end{equation}
where $T^a$ are the generators of the corresponding gauge group.
The non-Abelian aether term was generated for this theory in \cite{ouraether}, where it was shown to be finite and ambiguous, similarly to its Abelian analogue, cf. \cite{Gomes:2009ch}.

At the same time, the generation of the triple and quartic terms in the Yang-Mills action using only minimal couplings is a nontrivial problem. This is the aim we pursue in this paper.
We start with the following action of the spinor coupled to the non-Abelian gauge field:
\bea
S=\int d^4x \bar{\psi}^i \left(i\ds\delta^{ij}-e\As^a(T^a)^{ij}-m\delta^{ij}-\bs\gamma_5\delta^{ij}
\right)\psi^j.
\eea
Unlike the nonminimal coupling considered in \cite{ouraether}, this action involves the minimal coupling only, with the corresponding coupling constant being dimensionless. As a result, this theory is all-loop renormalizable.

Within our study, our aim consists in computing the one-loop effective action presented by the following fermionic determinant
\bea
\Gamma^{(1)}=i {\rm Tr}\ln (i\ds-e\As^aT^a-m-\bs\gamma_5),
\eea
where $\As^a=\gamma^{\mu}A_{\mu}^a$, since our vector field is Lie-algebra valued.
Expanding the fermionic determinant up to the fourth order in external fields, we find (here $\As=\As^aT^a$)
\bea
-i\Gamma^{(1)}&=&-\frac{e^2}{2}{\rm Tr}\As\frac{1}{i\ds-m-\bs\gamma_5}\As\frac{1}{i\ds-m-\bs\gamma_5}+\nonumber\\
&+&\frac{e^3}{3}{\rm Tr}\As\frac{1}{i\ds-m-\bs\gamma_5}\As\frac{1}{i\ds-m-\bs\gamma_5}\As\frac{1}{i\ds-m-\bs\gamma_5}-\nonumber\\
&-&\frac{e^4}{4}{\rm Tr}\As\frac{1}{i\ds-m-\bs\gamma_5}\As\frac{1}{i\ds-m-\bs\gamma_5}\As\frac{1}{i\ds-m-\bs\gamma_5}\As\frac{1}{i\ds-m-\bs\gamma_5}.
\eea
To proceed with this calculation, one can use the exact propagator of the spinor whose form in the momentum space is given by \cite{Andri}:
\bea
S(k)=\frac{1}{\ks-m-\bs\gamma_5}=\frac{k^2+b^2-m^2+2(b\cdot k+m\bs)\gamma_5}{(k^2+b^2-m^2)^2-4[(b\cdot k)^2-m^2b^2]}(\ks+m+\bs\gamma_5).
\eea
However, within our purposes it is more convenient to use the usual propagator of $\psi$, that is, $<\bar{\psi}(-p)\psi(p)>=\frac{1}{\ps-m}$, since we consider the contributions only up to the second order in $b_{\mu}$. We note that since the aether term which will be obtained from a minimal coupling is non-ambiguous, the results obtained with use either of the modified propagator or the simple one will be the same.

The result for the second order in $A_{\mu}$ can be obtained through the sum of contributions for the three diagrams, with two insertions of $\slashed{b}\gamma_{5}$, carrying out the expansion of the following expression:
\begin{eqnarray}
\Gamma_{2}^{(1)}=\frac{ie^2}{2}\mbox{Tr}\slashed{A}\frac{1}{i\slashed\partial-m-\slashed{b}\gamma_{5}}\slashed{A}\frac{1}{i\slashed\partial-m-\slashed{b}\gamma_{5}}.
\end{eqnarray}

In this case, the total contribution is given by
\begin{eqnarray}\label{a5}
\Gamma_{2}^{(1)} &=& \frac{e^2}{2}\Big[\int\frac{d^4 p}{(2\pi)^4}\frac{2\mbox{tr}[\gamma^{\mu}(\slashed{p}+m)\gamma^{\nu}(\slashed{p}+\slashed{k}+m)\slashed{b}\gamma_{5}(\slashed{p}+\slashed{k}+m)\slashed{b}\gamma_{5}(\slashed{p}+\slashed{k}+m)]}{(p^2-m^2)^4}\nonumber\\
&+& \int\frac{d^4 p}{(2\pi)^4}\frac{\mbox{tr}[\gamma^{\mu}(\slashed{p}+m)\slashed{b}\gamma_{5}(\slashed{p}+m)\gamma^{\nu}(\slashed{p}+\slashed{k}+m)\slashed{b}\gamma_{5}(\slashed{p}+\slashed{k}+m)]}{(p^2-m^2)^4}\\
&+& \int\frac{d^4 p}{(2\pi)^4}\frac{\mbox{tr}[\gamma^{\mu}(\slashed{p}+m)\slashed{b}\gamma_{5}(\slashed{p}+m)\slashed{b}\gamma_{5}(\slashed{p}+m)\gamma^{\nu}(\slashed{p}+\slashed{k}+m)]}{(p^2-m^2)^4}\Big]A_{\mu}^{a}(-k)A_{\nu}^{b}(k)\mbox{tr}(T^{a}T^{b}).\nonumber
\end{eqnarray}

The result represents itself as a direct generalization of the Abelian contribution (see f.e. \cite{Scarp}) and looks like
\begin{eqnarray}
\label{2point}
\Gamma^{(1)}_2=-\frac{e^2}{6\pi^2m^2}b_{\mu}F^{\mu\nu a}_0b^{\lambda}F^b_{\lambda\nu 0}{\rm tr}(T^aT^b)=-\frac{\kappa e^2}{6\pi^2m^2}b_{\mu}F^{\mu\nu a}_0b^{\lambda}F^a_{\lambda\nu 0},
\end{eqnarray}
where $F^{\mu\nu a}_0=\pa^{\mu}A^{\nu a}-\pa^{\nu}A^{\mu a}$ is the Abelian part of the stress tensor, and the generators are normalized through the relation ${\rm tr}(T^aT^b)=\kappa\delta^{ab}$, with $\kappa\neq 0$ is a some real number defining normalization of the generators. 
This result can be represented as
\begin{eqnarray}
\Gamma^{(1)}_2=-\frac{\kappa e^2}{6\pi^2m^2}\Pi^{\lambda\rho\alpha\beta}\partial_{\lambda}A^a_{\rho}\partial_{\alpha}A_{\beta}^a,
\end{eqnarray} 
where
\begin{eqnarray}
\label{tensor}
\Pi^{\lambda\rho\alpha\beta}=\eta^{\rho\beta}b^{\lambda}b^{\alpha}-\eta^{\rho\alpha}b^{\lambda}b^{\beta}-\eta^{\lambda\beta}b^{\rho}b^{\alpha}+
\eta^{\lambda\alpha}b^{\rho}b^{\beta},
\end{eqnarray}
so that $\Pi^{\lambda\rho\alpha\beta}\partial_{\lambda}\partial_{\alpha}$ is a transversal operator.  Straightforward analysis of corresponding Feynman diagrams allows to show that just this operator arises when third-order  and fourth-order contributions are obtained.
We note that  these, higher-order contributions are essentially non-Abelian, being absent in the $U(1)$ case.

\section{Third-order contribution}

Our starting point is the three-point contribution to the effective action of the gauge field $A_{\mu}^{a}$  given by
\bea
\Gamma_{3}^{(1)}=\frac{ie^3}{3}\mbox{Tr}\slashed{A}\frac{1}{i\slashed\partial-m-\slashed{b}\gamma_{5}}\slashed{A}\frac{1}{i\slashed\partial-m-\slashed{b}\gamma_{5}}\slashed{A}\frac{1}{i\slashed\partial-m-\slashed{b}\gamma_{5}}.
\eea
Here, we use the usual propagator of the spinor field.

Throughout our computation we consider the terms up to the second order in $b^{\mu}$ (proportional to $b_{\mu}b^{\lambda}$, but we disregard all terms proportional to $b^2$ since they yield only Lorentz-invariant contributions). The terms of the second order in $b_{\mu}$ are given by Fig. 1.
\begin{figure}[htp!]
\includegraphics[scale=0.5]{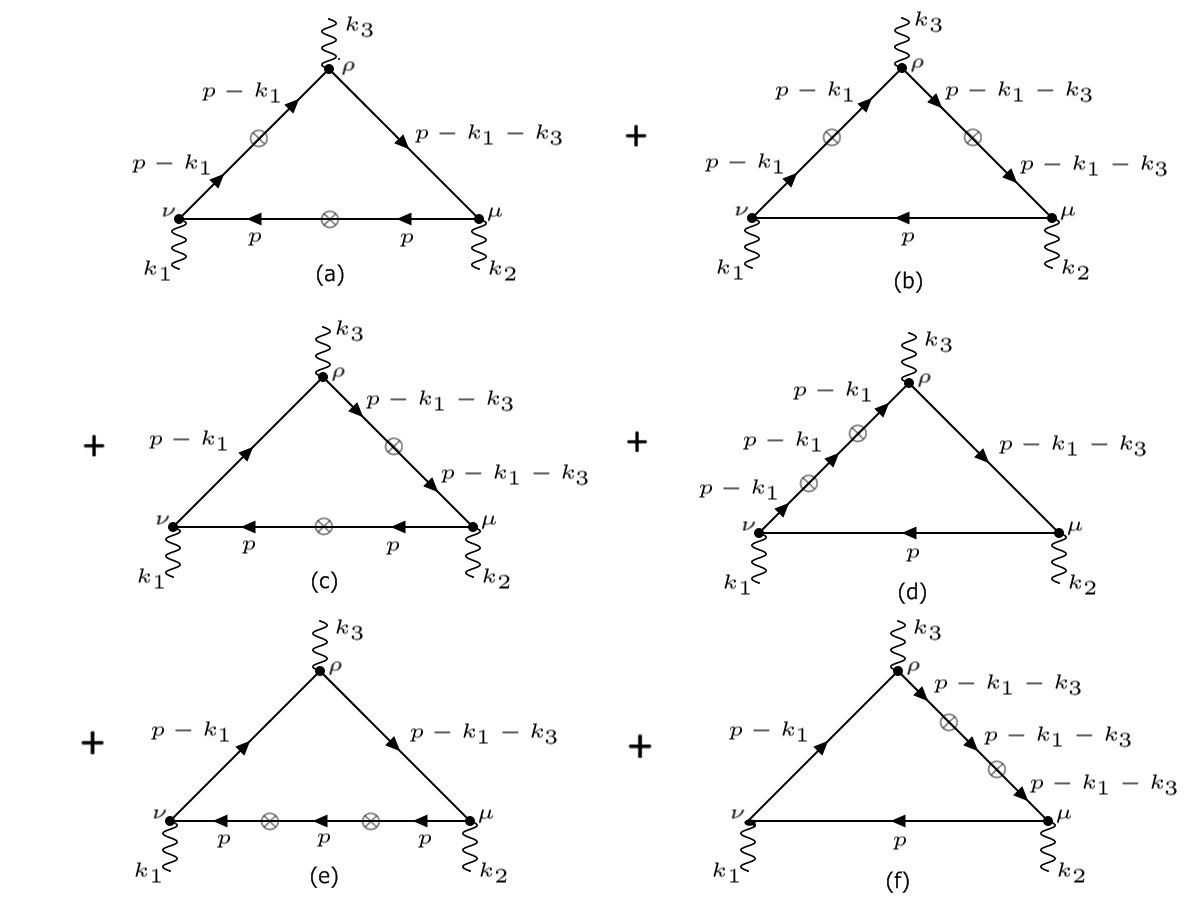}
\caption{Contributions of the third order in the external fields, with two insertions of $\slashed{b}\gamma_{5}$.}
\end{figure}

We note that the algebraic factors accompanying our quantum corrections are the same as in the usual Lorentz-invariant Yang-Mills theory since $\bs\gamma_5$ insertion is proportional to the Kronecker symbol $\delta^{ij}$ in the isotopic space (for a general discussion of quantum aspects of the Yang-Mills theory, see the classic papers \cite{classic}). These factors will yield first order in structure constants $f^{abc}$ for the three-point function, and the second order for the four-point function.

Carrying different contractions in a manner similar to calculations of higher-point functions presented in \cite{classic,PS}, we find that,
explicitly, the contribution (a) looks like
\begin{eqnarray}\label{a0}
\Gamma_{3,a}^{(1)}&=& \frac{e^3}{6}\int\frac{d^4 p}{(2\pi)^4}\mbox{tr}\left[\gamma^{\mu}
\frac{1}{\slashed{p}-m}
\slashed{b}\gamma_{5}\frac{1}{\slashed{p}-m}\gamma^{\nu}\frac{1}{\slashed{p}-\slashed{k_{1}}-m}\slashed{b}\gamma_{5}\frac{1}{\slashed{p}-\slashed{k_{1}}-m}\gamma^{\rho}\right.\times\nonumber\\&\times& \left.\frac{1}{\slashed{p}-\slashed{k_{1}}-\slashed{k_{3}}-m}\right]
A_{\mu}^{a}(k_1)A_{\nu}^{b}(k_2)A_{\rho}^{c}(k_3)\mbox{tr}(T^{a}[T^{b},T^{c}]),
\end{eqnarray}
where $k_1+k_2+k_3=0$.

We expand the propagators up to the first order in external momenta $k_{1}$ and $k_{3}$, and rewrite this contribution in the form:
\begin{eqnarray}\label{a1}
\Gamma_{3,a}^{(1)} &=& \frac{e^3}{3}{\rm tr}\int\frac{d^4 p}{(2\pi)^4}\Big[\frac{[\gamma^{\mu}(\slashed{p}+m)\slashed{b}\gamma_{5}(\slashed{p}+m)\gamma^{\nu}(\slashed{p}+m)\slashed{k}_1(\slashed{p}+m)\slashed{b}\gamma_{5}(\slashed{p}+m)\gamma^{\rho}(\slashed{p}+m)]}{(p^2-m^2)^6}\nonumber\\
&+&\frac{[\gamma^{\mu}(\slashed{p}+m)\slashed{b}\gamma_{5}(\slashed{p}+m)\gamma^{\nu}(\slashed{p}+m)\slashed{b}\gamma_{5}(\slashed{p}+m)\slashed{k}_1(\slashed{p}+m)\gamma^{\rho}(\slashed{p}+m)]}{(p^2-m^2)^6}\nonumber\\
&+& \frac{[\gamma^{\mu}(\slashed{p}+m)\slashed{b}\gamma_{5}(\slashed{p}+m)\gamma^{\nu}(\slashed{p}+m)\slashed{b}\gamma_{5}(\slashed{p}+m)\gamma^{\rho}(\slashed{p}+m)(\slashed{k}_1+\slashed{k}_3)(\slashed{p}+m)]}{(p^2-m^2)^6}
\Big]\times\nonumber\\&\times&
A_{\mu}^{a}(k_1)A_{\nu}^{b}(k_2)A_{\rho}^{c}(k_3)\mbox{tr}(T^{a}[T^{b},T^{c}[).
\end{eqnarray}

The contributions (a), (b) and (c) yield equal results  by symmetry reasons. 

On the other hand, for (d), (e) and (f) contributions, we proceed in a similar way. The (d) contribution looks like
\begin{eqnarray}\label{a01}
\Gamma_{3,d}^{(1)}&=& \frac{e^3}{6}{\rm tr}\int\frac{d^4 p}{(2\pi)^4}\left[\gamma^{\mu}
\frac{1}{\slashed{p}-m}
\gamma^{\nu}\frac{1}{\slashed{p}-\slashed{k_{1}}-m}\slashed{b}\gamma_{5}\frac{1}{\slashed{p}-\slashed{k}_1-m}\slashed{b}\gamma_{5}\frac{1}{\slashed{p}-\slashed{k_{1}}-m}\gamma^{\rho}\right.\times\nonumber\\&\times& \left.\frac{1}{\slashed{p}-\slashed{k_{1}}-\slashed{k_{3}}-m}\right]
A_{\mu}^{a}(k_1)A_{\nu}^{b}(k_2)A_{\rho}^{c}(k_3)\mbox{tr}(T^{a}[T^{b},T^{c}]),
\end{eqnarray}
After expansion in external momenta and keeping only the first-order terms in this expansion, one finds
\begin{eqnarray}\label{a4}
\Gamma_{3}^{(d)} &=& \frac{e^3}{3}{\rm tr}\Big[\int\frac{d^4 p}{(2\pi)^4}\frac{
\gamma^{\mu}(\slashed{p}+m)\gamma^{\nu}(\slashed{p}+m)\slashed{k}_1(\slashed{p}+m)\slashed{b}\gamma_5(\slashed{p}+m)\slashed{b}\gamma_5(\slashed{p}+m)\gamma^{\rho}(\slashed{p}+m)
}{(p^2-m^2)^6}\nonumber\\
&+& \int\frac{d^4 p}{(2\pi)^4}\frac{
\gamma^{\mu}(\slashed{p}+m)\gamma^{\nu}(\slashed{p}+m)\slashed{b}\gamma_5(\slashed{p}+m)\slashed{k}_1(\slashed{p}+m)\slashed{b}\gamma_5(\slashed{p}+m)\gamma^{\rho}(\slashed{p}+m)
}{(p^2-m^2)^6}\\
&+& \int\frac{d^4 p}{(2\pi)^4}\frac{
\gamma^{\mu}(\slashed{p}+m)\gamma^{\nu}(\slashed{p}+m)\slashed{b}\gamma_5(\slashed{p}+m)\slashed{b}\gamma_5(\slashed{p}+m)
\slashed{k}_1(\slashed{p}+m)\gamma^{\rho}(\slashed{p}+m)
}{(p^2-m^2)^6}
\nonumber\\
&+& \int\frac{d^4 p}{(2\pi)^4}\frac{
\gamma^{\mu}(\slashed{p}+m)\gamma^{\nu}(\slashed{p}+m)\slashed{b}\gamma_5(\slashed{p}+m)\slashed{b}\gamma_5(\slashed{p}+m)
\gamma^{\rho}(\slashed{p}+m)(\slashed{k}_1+\slashed{k}_3)(\slashed{p}+m)
}{(p^2-m^2)^6}
\Big]\times\nonumber\\&\times&A_{\mu}^{a}(k_1)A_{\nu}^{b}(k_2)A_{\rho}^{c}(k_3)
\mbox{tr}(T^{a}[T^{b},T^{c}]).\nonumber
\end{eqnarray}


The diagrams (d), (e) and (f) also yield equal results by symmetry reasons. Due to the presence of the commutators, all contributions turn out to be proportional to the first order in $f^{abc}$, just as in the Lorentz-invariant case \cite{classic}.

The complete result for the three-point function is a sum of results for (a) -- (f) diagrams given by Fig. 1. Through straightforward comparison with contributions to the two-point function, it is shown to be proportional to the same tensor $\Pi^{\lambda\rho\alpha\beta}$ (\ref{tensor}) arising within the calculation of the two-point function, explicitly,
\begin{eqnarray}
\Gamma^{(1)}_3=\frac{\kappa e^2}{6\pi^2m^2}f^{abc}\Pi^{\lambda\rho\alpha\beta}\partial_{\lambda}A^a_{\rho}A^b_{\alpha}A_{\beta}^c,
\end{eqnarray} 
or, as is the same,
\begin{eqnarray}
\label{3point}
\Gamma_{3}^{(1)} = \frac{\kappa e^3}{3\pi^{2}m^{2}}b_{\mu}f^{abc}F_{0}^{\mu\nu a}b^{\lambda}A_{\lambda}^{b}A_{\nu}^{c}.
\end{eqnarray}
We note that the constant factor accompanying this term is appropriate to form the gauge covariant expression for the non-Abelian gauge invariant action.


\section{Fourth-order contribution}

Now, we turn to the fourth-order contribution. It is given by
\bea
\Gamma_{4}^{(1)}=-\frac{e^4}{4}\mbox{Tr}\slashed{A}\frac{1}{i\slashed\partial-m-\slashed{b}\gamma_{5}}\slashed{A}\frac{1}{i\slashed\partial-m-\slashed{b}\gamma_{5}}\slashed{A}\frac{1}{i\slashed\partial-m-\slashed{b}\gamma_{5}}\slashed{A}\frac{1}{i\slashed\partial-m-\slashed{b}\gamma_{5}}.
\eea

The relevant contribution is presented by the Feynman diagram depicted at Fig. 2.

\begin{figure}[htbp]
\includegraphics[scale=0.3]{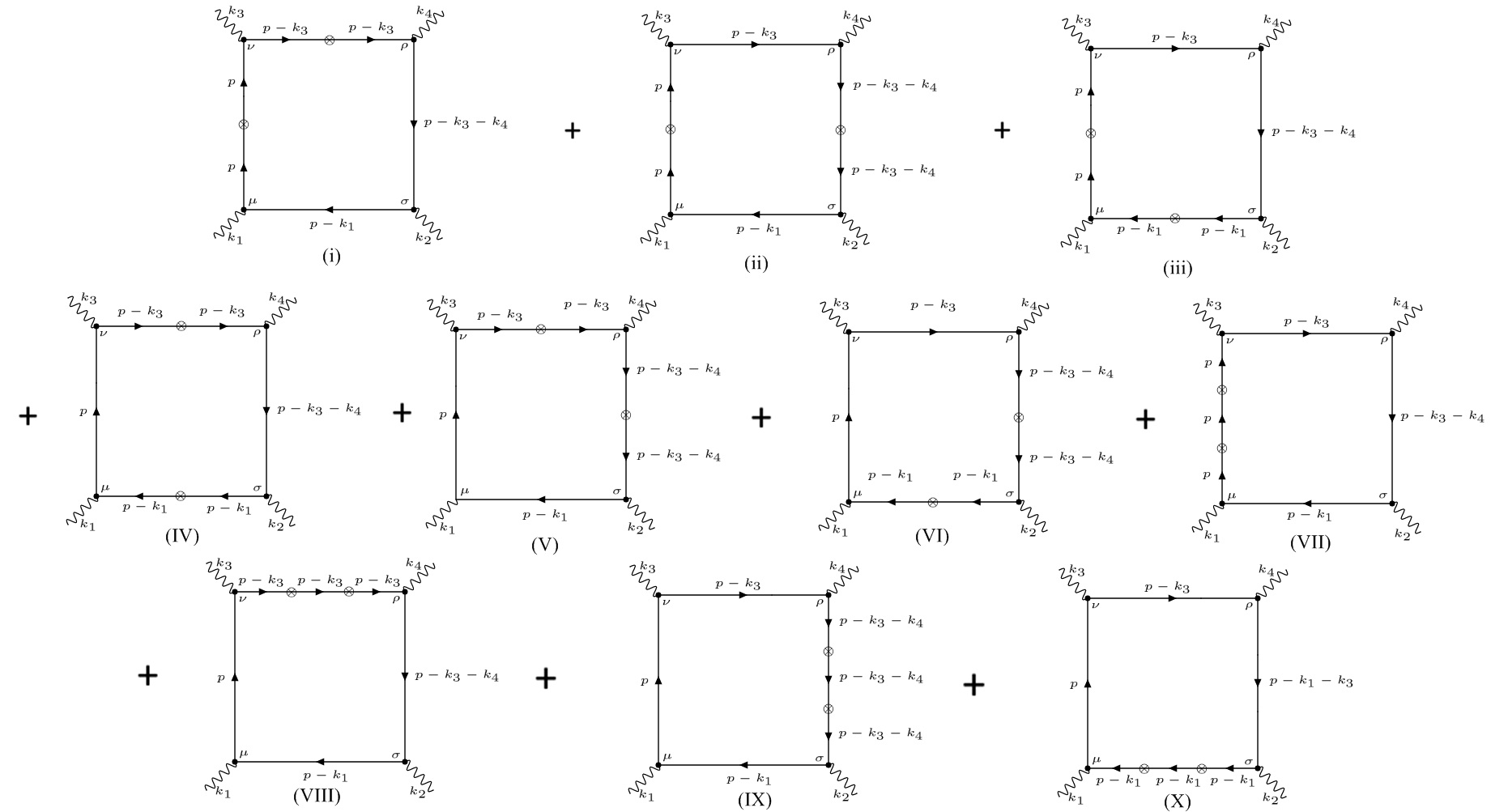}
\caption{Contribution of the fourth order in the external fields, with two insertions of $\slashed{b}\gamma_{5}$.}
\end{figure}

Explicitly, considering the cycle $(I)\,-\,(III)\,-\,(V)\,-\,(VI)$, with each contribution is similar to the ($I$), we have for $I$ contribution:
\begin{eqnarray}
\label{gamma4}
\Gamma_4^{(I)}&=& -\frac{e^4}{4}\int\frac{d^4 p}{(2\pi)^4}\mbox{tr}\left[\gamma^{\mu}\frac{(\slashed{p}+m)}{p^{2}-m^{2}}\slashed{b}\gamma_{5}\frac{(\slashed{p}+m)}{p^{2}-m^{2}}\gamma^{\nu}\frac{(\slashed{p}-\slashed{k}_{3}+m)}{(p-k_{3})^{2}-m^{2}}\slashed{b}\gamma_{5}\frac{(\slashed{p}-\slashed{k}_{3}+m)}{(p-k_{3})^{2}-m^{2}}\right.\times\nonumber\\&\times& \left.\gamma^{\rho}\frac{(\slashed{p}-\slashed{k}_{3}-\slashed{k}_{4}+m)}{(p-k_{3}-k_{4})^{2}-m^{2}}\gamma^{\lambda}\frac{(\slashed{p}-\slashed{k}_{1}+m)}{(p-k_{1})^{2}-m^{2}}\right]
A_{\mu}^{a}A_{\nu}^{b}A_{\lambda}^{c}A_{\rho}^d\mbox{tr}([T^{a},T^{b}][T^{c},T^{d}]).
\end{eqnarray}
Here, the commutators again arise due to various manners of carrying out the contractions, as in \cite{PS}.
Now, to form the contribution to the Yang-Mills-aether term, only zero order in momenta must be kept, so,
\begin{eqnarray}
\Gamma_{4}^{(I)} &=& -\frac{e^4}{4}\int\frac{d^4 p}{(2\pi)^4}\mbox{tr}\left[\frac{\gamma^{\mu}(\slashed{p}+m)\slashed{b}\gamma_{5}(\slashed{p}+m)\gamma^{\nu}(\slashed{p}+m)\slashed{b}\gamma_{5}(\slashed{p}+m)\gamma^{\rho}(\slashed{p}+m)\gamma^{\lambda}(\slashed{p}+m)}{(p^2-m^2)^6}\right]\times\nonumber\\&\times&
A_{\mu}^{a}A_{\nu}^{b}A_{\lambda}^{c}A_{\rho}^d\mbox{tr}([T^{a},T^{b}][T^{c},T^{d}]).
\end{eqnarray}

The diagrams $(II),(IV)$ are equal. Explicitly,
the contribution ($IV$) is 
\begin{eqnarray}
\label{gamma5}
\Gamma_4^{(IV)}&=& -\frac{e^4}{4}\int\frac{d^4 p}{(2\pi)^4}\mbox{tr}\left[\gamma^{\mu}\frac{(\slashed{p}+m)}{p^{2}-m^{2}}\gamma^{\nu}\frac{(\slashed{p}-\slashed{k}_3+m)}{(p-k_3)^{2}-m^{2}}\slashed{b}\gamma_{5}\frac{(\slashed{p}-\slashed{k}_3+m)}{(p-k_{3})^{2}-m^{2}}\right.\times\nonumber\\&\times& \left. \gamma^{\lambda}\frac{(\slashed{p}-\slashed{k}_{3}-\slashed{k}_{4}+m)}{(p-k_{3}-k_{4})^{2}-m^{2}}\gamma^{\rho}\frac{(\slashed{p}-\slashed{k}_{1}+m)}{(p-k_{1})^{2}-m^{2}}\slashed{b}\gamma_{5}\frac{(\slashed{p}-\slashed{k}_{1}+m)}{(p-k_{1})^{2}-m^{2}}\right]\times\nonumber\\&\times&
A_{\mu}^{a}A_{\nu}^{b}A_{\lambda}^{c}A_{\rho}^d\mbox{tr}([T^{a},T^{b}][T^{c},T^{d}]).
\end{eqnarray}
Keeping again only zero order in momenta, we have
\begin{eqnarray}
\Gamma_{4}^{(IV)} &=& -\frac{e^4}{4}\int\frac{d^4 p}{(2\pi)^4}\mbox{tr}\left[\frac{\gamma^{\mu}(\slashed{p}+m)\gamma^{\nu}(\slashed{p}+m)\slashed{b}\gamma_{5}(\slashed{p}+m)\gamma^{\lambda}(\slashed{p}+m)\gamma^{\rho}(\slashed{p}+m)\slashed{b}\gamma_{5}(\slashed{p}+m)}{(p^2-m^2)^6}\right]\times\nonumber\\&\times&
A_{\mu}^{a}A_{\nu}^{b}A_{\lambda}^{c}A_{\rho}^d\mbox{tr}([T^{a},T^{b}][T^{c},T^{d}]). 
\end{eqnarray}

The diagrams $(VII)\,-\,(VIII)\,-\,(IX)-\,(X)$ yield the same contributions.
Explicitly, the contribution ($X$) is 
\begin{eqnarray}
\label{gamma6}
\Gamma_4^{(X)}&=& -\frac{e^4}{4}\int\frac{d^4 p}{(2\pi)^4}\mbox{tr}\left[\gamma^{\mu}\frac{(\slashed{p}+m)}{p^{2}-m^{2}}\gamma^{\nu}\frac{(\slashed{p}-\slashed{k_3}+m)}{(p-k_3)^{2}-m^{2}}\gamma^{\lambda}\frac{(\slashed{p}-\slashed{k}_3-\slashed{k}_4+m)}{(p-k_{3}-k_{4})^{2}-m^{2}}\right.\times\nonumber\\&\times& \left. \gamma^{\rho}\frac{(\slashed{p}-\slashed{k}_{1}+m)}{(p-k_{1})^{2}-m^{2}}\slashed{b}\gamma_{5}\frac{(\slashed{p}-\slashed{k}_{1}+m)}{(p-k_{1})^{2}-m^{2}}\slashed{b}\gamma_{5}\frac{(\slashed{p}-\slashed{k}_{1}+m)}{(p-k_{1})^{2}-m^{2}}\right]\times\nonumber\\&\times&
A_{\mu}^{a}A_{\nu}^{b}A_{\lambda}^{c}A_{\rho}^d\mbox{tr}([T^{a},T^{b}][T^{c},T^{d}]).
\end{eqnarray}
Again, keeping only zero order in momenta, we have
\begin{eqnarray}
\Gamma_{4}^{(X)} &=& -\frac{e^4}{4}\int\frac{d^4 p}{(2\pi)^4}\mbox{tr}\left[\frac{\gamma^{\mu}(\slashed{p}+m)\gamma^{\nu}(\slashed{p}+m)\gamma^{\lambda}(\slashed{p}+m)\gamma^{\rho}(\slashed{p}+m)\slashed{b}\gamma_{5}(\slashed{p}+m)\slashed{b}\gamma_{5}(\slashed{p}+m)}{(p^2-m^2)^6}\right]\times\nonumber\\&\times&
A_{\mu}^{a}A_{\nu}^{b}A_{\lambda}^{c}A_{\rho}^d\mbox{tr}([T^{a},T^{b}][T^{c},T^{d}]).
\end{eqnarray}

Calculating all traces and using the integrals listed in the appendix \ref{listint}, we show that the result is proportional to the second order in structure constants, as it must be by the gauge symmetry reasons, and to the operator $\Pi^{\lambda\rho\alpha\beta}$, and has the following form
\begin{eqnarray}
\Gamma^{(1)}_4=-\frac{\kappa e^4}{24\pi^2m^2}\Pi^{\lambda\rho\alpha\beta}f^{abm}f^{cdm}A_{\lambda}^aA_{\rho}^b A_{\alpha}^{c}A_{\beta}^{d},
\end{eqnarray} 
or, as is the same,
\begin{eqnarray}
\label{4point}
\Gamma_{4}^{(1)} = -\kappa\frac{e^4}{6\pi^{2}m^{2}}f^{abm}f^{cdm}b_{\mu}A^{\mu a}A^{\nu b}b^{\lambda}A_{\lambda}^{c}A_{\nu}^{d}.
\end{eqnarray}
The sum of the expressions \eqref{2point}, \eqref{3point} and \eqref{4point} yields the desired result
\bea
\Gamma^{(1)}=-\kappa\frac{e^2}{6\pi^2m^2}b^{\mu}F_{\mu\nu}^ab_{\lambda}F^{\lambda\nu a},
\eea
where
\bea
F^{\mu\nu a}=\partial^{\mu}A^{\nu a}-\partial^{\nu}A^{\mu a}-ef^{abc}A^{\mu b}A^{\nu c}
\eea
is the non-Abelian stress tensor. 
Thus, we conclude that we have succeeded to generate the non-Abelian aether-like term.
 
 \section{Summary}
 
In this paper,  we have performed the generation of  non-Abelian aether-like term. Our starting point was the theory of Dirac spinor minimally coupled to the non-Abelian gauge field, where the Lorentz symmetry breaking has been introduced through the usual $\bs\gamma_5$ term, that is, the same model considered in \cite{YMCS} and representing itself as a straightforward non-Abelian generalization of the model used within studies of a Lorentz-breaking extension of QED.
 
 The main significance of our result consists in the fact that while up to now, most of the papers devoted to study of perturbative aspects of Lorentz-breaking theories were concentrated either on quadratic finite contributions (for a review on finite corrections see \cite{ourreview}) or on renormalization of coupling vertices (see f.e. \cite{KosPic}), here we performed perturbative generation of a finite non-Abelian, fourth-order contribution. We note that our result is valid for an arbitrary gauge group. It is the second example of generation of a non-Abelian term carried out with use of only minimal couplings, after \cite{YMCS}, ever realized.  Actually, our result is the next-order contribution to the expansion of the effective action of the non-Abelian gauge field coupled to fermions, after the non-Abelian CFJ term \cite{YMCS}. The advantage of our approach in comparison with \cite{ouraether} consists in the fact that unlike the scheme presented in \cite{ouraether}, the calculation performed in this paper is carried out on the base of a minimal coupling, being hence explicitly superficially finite and hence ambiguity-free. Thus, we have confirmed that the non-Abelian Lorentz-breaking terms can arise as quantum corrections in a some fundamental theory, as it follows from the concept of emergent dynamics.
 
As it was discussed in \cite{ouraether}, the impact of the aether term should be small in comparison with the usual Yang-Mills term since the Lorentz-breaking parameters are very small. It is natural to expect that there will be only small modifications in qualitative description of the confinement generated by the aether term. Nevertheless we note that there are many issues related to the non-Abelian aether term within the confinement problem and other contexts which need to be studied. We expect to perform such studies in our next papers.

{\bf Acknowledgements.} Authors are grateful to J. C. Rocha for important discussions. This work was partially supported by Conselho
Nacional de Desenvolvimento Cient\'{\i}fico e Tecnol\'{o}gico (CNPq). The work by A. Yu. P. has been supported by the
CNPq project No. 303783/2015-0.

\appendix
\section{List of integrals}
\label{listint}
Here we list the integrals used to perform our computations:
\begin{eqnarray}
\int\frac{d^4 p}{(2\pi)^4}\frac{1}{(p^2-m^2)^5}&=&-\frac{i}{192\pi^{2}m^{6}};\\
\int\frac{d^4 p}{(2\pi)^4}\frac{p^{2}}{(p^2-m^2)^5}&=&\frac{i}{192\pi^{2}m^{4}};\\
\int\frac{d^4 p}{(2\pi)^4}\frac{p^{\mu}p^{\nu}}{(p^2-m^2)^5}&=&\frac{ig^{\mu\nu}}{768\pi^{2}m^{4}};\\
\int\frac{d^4 p}{(2\pi)^4}\frac{p^{4}}{(p^2-m^2)^5}&=&-\frac{i}{64\pi^{2}m^{2}};\\
\int\frac{d^4 p}{(2\pi)^4}\frac{1}{(p^2-m^2)^6}&=&-\frac{i}{320\pi^{2}m^{8}};\\
\int\frac{d^4 p}{(2\pi)^4}\frac{p^{2}}{(p^2-m^2)^6}&=&-\frac{i}{480\pi^{2}m^{6}};\\
\int\frac{d^4 p}{(2\pi)^4}\frac{p^{\mu}p^{\nu}}{(p^2-m^2)^6}&=&-\frac{ig^{\mu\nu}}{1920\pi^{2}m^{6}};\\
\int\frac{d^4 p}{(2\pi)^4}\frac{p^{4}}{(p^2-m^2)^6}&=&\frac{i}{320\pi^{2}m^{4}};\\
\int\frac{d^4 p}{(2\pi)^4}\frac{1}{(p^2-m^2)^7}&=&-\frac{i}{480\pi^{2}m^{10}};\\
\int\frac{d^4 p}{(2\pi)^4}\frac{p^{2}}{(p^2-m^2)^7}&=&\frac{i}{960\pi^{2}m^{8}}\\
\int\frac{d^4 p}{(2\pi)^4}\frac{p^{\mu}p^{\nu}}{(p^2-m^2)^7}&=&\frac{ig^{\mu\nu}}{3840\pi^{2}m^{8}};\\
\int\frac{d^4 p}{(2\pi)^4}\frac{p^{4}}{(p^2-m^2)^7}&=&-\frac{i}{960\pi^{2}m^{6}};\\
\int\frac{d^4 p}{(2\pi)^4}\frac{p^{\mu}p^{\nu}p^{\rho}p^{\sigma}}{(p^2-m^2)^5}&=&-\frac{i}{1536\pi^{2}m^{2}}\left(g^{\mu\nu}g^{\rho\sigma}+g^{\mu\rho}g^{\nu\sigma}+g^{\mu\sigma}g^{\nu\rho}\right);\\
\int\frac{d^4 p}{(2\pi)^4}\frac{p^{\mu}p^{\nu}p^{\rho}p^{\sigma}}{(p^2-m^2)^6}&=&\frac{i}{7680\pi^{2}m^{4}}\left(g^{\mu\nu}g^{\rho\sigma}+g^{\mu\rho}g^{\nu\sigma}+g^{\mu\sigma}g^{\nu\rho}\right);\\
\int\frac{d^4 p}{(2\pi)^4}\frac{p^{\mu}p^{\nu}p^{\rho}p^{\sigma}}{(p^2-m^2)^7}&=&-\frac{i}{23040\pi^{2}m^{6}}\left(g^{\mu\nu}g^{\rho\sigma}+g^{\mu\rho}g^{\nu\sigma}+g^{\mu\sigma}g^{\nu\rho}\right).
\end{eqnarray}


\end{document}